\documentclass [prb,preprint,showpacs,superscriptaddress]{revtex4}
\usepackage{graphicx}

\begin{document}

\title{Changes in the surface electronic structure upon martensitic transformation in TiNi and TiPd}

\author{S.E. Kulkova}
\affiliation{Institute of Strength Physics and Materials Science of the Russian Academy of Sciences, 634021, Tomsk, Russia}

\author{V.E. Egorushkin}
\affiliation{Tomsk State University, 634050, Tomsk, Russia}

\author{J.S. Kim, G. Lee and Y.M. Koo}
\affiliation{Department of Physics, Pohang University of Science and Technology, Pohang 790-784, Republic of Korea}

\date{\today}

\begin{abstract}
The electronic structure of some low index surfaces for martensitic 
$B$19'-TiNi and $B$19-TiPd were investigated using the full-potential linearized 
augmented plane wave method. The alteration of the electronic structure upon 
martensitic transformation on the surface were analyzed. According to the 
surface calculations only insignificant changes in the electronic structure 
near the Fermi level were found as for bulk ground state. The formation of 
oxides top layers is also discussed. It is shown that the influence of TiNi 
substrate is insignificant when three oxides layers are formed.
\end{abstract}


\maketitle

The electronic structure (ES) of the intermetallic TiNi and TiPd alloys were 
on the focus of a number of
works\cite{Papacon1971,Zhao1989,Kulkova1991,Bihlmayer1992,Sanati1998},
due to their unique mechanical 
properties and technological importance for engineering and medicine. The 
unusual mechanical properties such as shape memory effect (SME) are 
connected with martensitic transformations (MT's), which involve the 
displacements of atoms within the lattice together with the changes of its 
shape. This is a type of diffusionless structural phase transformation. The 
MT critical temperature for TiPd is around 840 K that much higher than that 
for TiNi ($\sim $333 K). On cooling $B$2-TiNi transforms martensitically to 
monoclinic $B$19' structure whereas TiPd goes to the orthorhombic $B$19 phase. The 
martensitic transformation sequence in TiNi strongly depends on the 
composition of the alloy, the doping with other metals, and prior 
thermochemical history. The results obtained in
Refs. \onlinecite{Papacon1971,Zhao1989,Kulkova1991,Bihlmayer1992,Sanati1998} and the references 
therein provide an insight into the mechanisms of the phase stability in 
Ti-based alloys. However, the surface ES of Ti-based alloys is less 
investigated\cite{Canto1999,Lee2001,Koroteev2001} in comparison with the bulk ground state. Many 
surface phenomena are directly related to the surface electronic structure. 
To the best of our knowledge, fundamental informations about the surface ES 
of $B$2-TiPd or matrensitic phases of both alloys have not been reported. Since 
TiNi has found widespread application in medicine in last decade, the 
investigation of its interactions with different adsorbates, protein or bone 
tissue is very important task. In order to explain complex processes on the 
surface at the microscopic level we need to understand the electronic 
structure of clean Ti-based alloys surfaces first of all. In our previous 
paper\cite{Lee2001} we have performed as a preliminary step before these 
calculations, a comparative study of the surface ES in a series of 
$B$2-Ti\textit{Me} alloys (\textit{Me}=Ni, Co, Fe). We have found the increase of density of states 
(DOS) at the Fermi level $N(E_{F})$ in the surface layers for all 
Ti-terminated $B$2-Ti\textit{Me} (001) films and also in $B$2-TiNi and TiFe (110) surface 
layers. The obtained results indicate that the Ti-atoms at the surface 
possess a high chemical reactivity. Since the surface Fe states are dominant 
at the Fermi level for both surfaces it is possible to conclude that Fe 
states are also highly reactive. However, it is necessary to take into 
account the magnetic ordering for the Fe-terminated (001) surface. The 
authors of Ref. \onlinecite{Shabal1985} inferred that the structural instability of the bulk 
Ti-based alloys is due to the larger Ti states contribution in 
$N(E_{F})$ in comparison with that of \textit{Me}. Our results\cite{Lee2001} showed substantial 
ES changes in the surface layer for both low index surfaces. These ES 
changes can alter MT's with respect to the martensitic transformation 
temperature as well as MT sequence on the surface. Thus MT's on the surface 
can be different from those in the bulk. In this paper we present the 
comparative study of surface ES for TiNi and TiPd martensitic phases.

Using the all-electron first principles technique (the Wien 97 
implementation of the full-potential linearized augmented plane wave (FLAPW) 
method\cite{Blaha1990} within local density approximation (LDA) for the 
exchange-correlation potential) we have investigated the surface ES of 
$B$19'-TiNi and $B$19-TiPd. The surface was simulated by repeated slabs separated 
by a vacuum region in one direction. The thickness of vacuum was chosen to 
be equal to the two lattice parameters of the bulk material. A five and 
seven atomic layer slabs were used for $B$2-TiNi and TiPd (110) and (001) 
surfaces simulation, whereas for $B$19'-TiNi and $B$19-TiPd surfaces a single 
slab, which consist of eight atomic layers, were used. The lateral lattice 
parameters were set to the experimental ones\cite{Kudoh1985,Dwight1965} for bulk alloys. Only 
the ideal martensitic structures were investigated but we estimated the 
relaxation effects for $B$2-TiNi and $B$2-TiPd (110). The interlayer distances in 
this case were optimized with Newton dynamics. The core states were treated 
in a non-relativistic fashion. The 3$s$-, 3$p$- states for TiNi and 4$s$-, 
4$d$-states for TiPd were treated as the valence band states. The multipole 
expansion of the crystal potential and electron density inside 
``muffin-tin'' spheres was cut at $l_{max}$=10. Non-spherical contributions 
to charge density and potential within spheres were considered up to 
$l_{max}$=4. In the interstitial region the charge density and crysral 
potential was expanded as a Fourier series with wave vectors up to 
$G_{max}$=10 a.u.$^{-1}$. The plane wave cuttof was 4 a.u.$^{-1}$. The 
density of states was calculated with 28-36 \textbf{k}-points grid in the 
irreducible part of two-dimensional Brillouin zone (2D BZ) depending on the 
considered surface. The calculated DOS was smoothed out by a Gaussian 
function with a width of 0.1 eV to suppress noise. Self-consistency was 
considered to have been achieved when the total energy variation from 
iteration to iteration did not exceed 10$^{-4}$ Ry. 

The low-temperature martensitic $B$19'-TiNi phase is formed from unit cell 
twice as large as that of the parent $B$2-phase with corresponding rotation of 
new cell around $B$2 [001] axis and also its shift along $B$2 [001] and [110] 
directions. The martensitic unit cell is compressed along $B$2 [001] and [110] 
directions but it is expanded along [$1\bar {1}0$]. Finally, both Ti and Ni 
atoms are shifted in the (110) plane, which is monoclinically distorted. 
Both Ti and Pd atoms are shifted in $B$2 [110] and [$1\bar {1}0$] directions 
but without monoclinic distortion. The atomic structure of $B$19'-TiNi (100), 
(010) and (001) surfaces are given in Fig. \ref{fig1}. The $B$19'-TiNi (001) surface has 
an oblique lattice with the same composition as the parent $B$2 (110) surface. 
The $B$19 and $B$19' (010) surface is rumpled with the Ni (Pd) atoms shifted by 
$\sim $0.5~(0.68) {\AA} with respect to Ti positions. Both $B$19'-TiNi (001) and 
(010) surfaces are derived from $B$2-TiNi (110) plane upon MT. The $B$19' (100) 
surface is derived from $B$2 (001) plane. In the last case the additional 
splitting within Ni and Ti layers is occurred (Fig. \ref{fig1}a). This leads to 
different variants in the surface termination. The splitting of the layers 
is consequence of the monoclinic distortion of the plane. The corresponding 
$B$19-TiPd (100) surface does not have such peculiarity. The ES result for the 
surface termination with two Ti top layers is given in Fig. \ref{fig2}a. Among three 
surface considered for martensitic $B$19'-TiNi, the (010) plane is less 
distorted in comparison with $B$2 (110) parent plane. It is the coherent 
boundary between both austenitic and martensitic phases. 

Differences in the surface electronic structure of both TiNi and TiPd alloys 
are reflected on the densities of states in Fig. \ref{fig2}, where the Fermi level is 
set to zero. For both austenitic and martensitic phases, the local DOS 
(LDOS) at the central layers display similar features to those of the bulk. 
We found that bulk region is correctly reproduced in all surface ES 
calculations. As seen from Fig. \ref{fig3} the LDOS curves obtained in bulk 
calculation for $B$19'-TiNi coincide well with those for central layers in 
surface calculations. This fact reveals that the perturbation of vacuum is 
well screened within the first few atomic layers from the surface. We can 
conclude that upon MT surface ES changes in the same way as in bulk 
alloys\cite{Kulkova1991} (a small decrease of $N(E_{F})$ in martensitic phase in 
comparison with austenitic one). Basically the redistribution of the states 
near $E_{F }$is observed. The $d$-bands of the alloys components become narrower 
in the surface layer. The change of ES is less for $B$19'-TiNi (001) surface 
despite the fact that it is most distorted plane. It was derived from the 
$B$2 (110) by following distortion: the change of the angle between two axes 
and the corresponding compression in one direction but the expansion in 
other one. In addition the Ti and Ni atoms are shifted in the plane with 
respect to their positions in the parent $B$2-phase. The Fermi level is 
dominated by Ti-states as in $B$2 (110). Moreover, it lies exactly in a small 
DOS peak that may be the indication of the instability of this surface. The 
same feature is found for $B$19-TiPd (001) surface but in this case the DOS 
peak is more broadened. 

One the possible mechanism of surface reconstruction is the displacement of 
Ni (Pd) atoms with respect to Ti atoms that leads to rumpled surface 
structure. As seen from Fig. \ref{fig1}b the idealized$ B$19'- and $B$19 (010) structures are 
rumpled ones. This plane is genetically derived from $B$2 ($1\bar {1}0)$ with 
slight compression in both directions upon MT. Initially in the $B$2 ($1\bar 
{1}0)$ plane the Ti atoms are already shifted with respect to Ni or Pd atoms 
during surface relaxation as follows from our calculations. Both Ti and 
Ni(Pd) atoms display a general downward movement. The obtained values of 
relaxation are $-$6.7 {\%}, $-$1.96 {\%} for Ti and Pd respectively in $B$2-TiPd 
(110) and $-$6.1 {\%}, $-$15.2 {\%} for Ti and Ni atoms in $B$2-TiNi (110). The 
striking feature of a sharp decrease of $N(E_{F})$ is found in the surface 
layer for $B$19'-TiNi (010). The Fermi level lies in the local minimum of DOS 
similarly to the bulk.\cite{Kulkova1991,Sanati1998} The shape of Ni LDOS changes substantially 
whereas the occupancies of the Ni-bonding orbitals are practically the same 
as for $B$19'-TiNi (001) surface. Moreover, the $B$19'-TiNi (010) surface has the 
lower total energy value in comparison with other cases. We infer that the 
$B$19'-TiNi (010) surface is the free energy surface and the growth of the 
crystal will be preferable in this plane. This is in good agreement with the 
experimental results.\cite{Gunter1992} We would like to emphasize that the 
displacements of atoms during relaxation in initial $B$2-TiNi (110) system are 
partially the same, which is realized upon MT. Thus the $B$2-TiNi (110) surface 
is most preferable for martensitic reconstruction. It is believed that the 
$B$2-TiNi (110) surface relaxation can stimulate MT. So the parent $B$2-Ti\textit{Me} (110) 
surface, where Me is 3$d$-5$d$ transition elements is partially prepared due to 
relaxation mechanism to future martensitic transformations. Nevertheless 
this feature cannot be the only driving force of the phase transformation. 
For example $B$2-TiFe (110) has the same rumpled surface structure but TiFe 
does not undergo MT. It is well known that TiFe has a very stable $B$2-phase in 
the TiFe-TiCo-TiNi-TiPd series. The ES peculiarities of the parent phase are 
very important in this context and the local effects can affect considerably 
MT with respect to martensitic temperature as well as martensite structure. 
A small amount of Fe ($\sim $1-2 {\%}) leads to essential decrease of 
MT.\cite{Shabal1985} The addition of Cu or Pd in $B$2-TiNi matrix leads to $B$2-$B$19 MT. The 
role of the interface will be investigated in more details in our 
forthcoming paper. Our results indicate that processes in the $B$2 (110) can be 
fundamental. At the present time it will be interesting to understand the 
influence of the alloying on the relaxation phenomena and to link them with 
the corresponding change of the MT temperature. The attempts to find the 
correlation between the changes in the MT temperature and bulk ES were 
recently made.\cite{Shabal1985,Cai1999} It is evident that if the $B$2 (110) surface can be 
prepared due to relaxation mechanism to future MT, it might lead to a 
decrease of the energy needed for this transformation.

In the case of $B$19'-TiNi (100) surface the presence of Ti-surface states near 
the Fermi level indicates the high chemical reactivity of this surface as 
well as parent $B$2-TiNi (001). The $B$19-TiPd (100) structure has a sharp peak at 
$E_{F }$but the Fermi level is slightly shifted from it for $B$19' structure due 
to additional splitting within Ni and Ti layers, which does not occur in the 
$B$19-TiPd (100) system. So, the splitting of layers can stabilize this 
structure at the surface. In both cases Ti atoms are highly reactivity on 
the surface. We simulate the decrease of the Ti concentration (additional 
calculation without one Ti surface layer was performed) but this effect does 
not lead to substantial DOS change near $E_{F}$. We analyzed also the DOS 
differences between surface and central layers in both austenitic and 
martensitic phase for all considered surfaces. The lowering of symmetry in 
martensitic phase leads to splitting of the surface states and their 
displacement along the energy scale. In general, the similar changes at the 
surfaces are observed in both alloys. The detailed analysis showed that the 
number of states and their orbital composition changes insignificantly upon 
MT. The alteration of the surface ES in TiPd upon MT is found be similar to 
TiNi. However, the DOS results for both $B$19-TiPd (010) and (100) surfaces 
indicate their instability. It seems that the monoclinic distortion is a 
consequence of the instability of the $B$19-TiNi (010) surface, as in the 
bulk.\cite{Kulkova1991} Actually the appearance of the monoclinic phase is due to the 
coexistence of the R-type\cite{Savushkin1991} and orthorhombic distortion. 

Thus, we can conclude that the mechanisms promoting the MT in the bulk can 
similarly give rise to that at the surface. In the bulk the atomic bonding 
can be changed with deviation from stoichiometry, defects, etc. However, 
even uniform change in the interface region due to impurity can lead to 
non-uniform displacements of the atoms along the direction, which is 
perpendicular to the considered plane. This can develop the mechanism, which 
was discussed above for the surface. The investigation of the influence of 
impurities on the interface behavior is in progress now.

It is assumed that a biocompatibility of TiNi is related to the presence of 
titanium oxides on the surface. Since TiNi-based alloys are used for bone, 
teeth implants and others medical applications, it is desirable to 
understand the different kind of the interactions on the metal oxide surface 
in that number with human tissue. The formation of the surface oxide layer 
could be also considered on $B$2-TiNi (110) surface. We simulated the TiO$_{2}$ 
(100) formation on the $B$2-TiNi (110) surface. Our slab consists of three 
layers of TiNi and four, six layers of oxide. It is evident that we used a 
very simplified model. The situation in the interface region can be very 
complicated and depends very much on the O coverage. The oxygen 
incorporation can induce a substantial distortion of the alloy lattice. In 
our case the lattice parameters differ slightly as they are 4.594 {\AA} and 2.98 {\AA}
for TiO$_{2}$ (100) and 4.264 {\AA} and 3.015 {\AA} for TiNi(110). The layer resolved 
DOS curves for TiO$_{2}$ on the TiNi substrate are given in Fig. \ref{fig4}a. We 
investigated also TiO$_{2}$ (100), (001) and (110) surfaces and found that 
our results are in consistent with previous ones reported in Refs. \onlinecite{Mackrodt1997,Vogtenhuber1994,Ramamoorty1994} and 
references therein. Since TiO$_{2}$ surfaces have been the subject of 
numerous works we will not describe the result here. It should be noted that 
basically the results were discussed in the earlier papers with respect to 
band gap states, which have been observed in the photoemission experiments 
as well as on oxygen deficient TiO$_{2}$(100) and (110) surfaces and also 
the surface reconstruction. It was shown that oxygen vacancies play a 
crucial role in the formation of the defects phases, the surface 
reconstructions and electronic properties of oxides layers. We did not found 
band gap states for stoichiometry composition in all considered surfaces. 
The obtained DOS's for clean TiO$_{2}$ (100) surface are presented in Fig. 
\ref{fig4}b. As seen from Fig. \ref{fig4} the influence of TiNi substrate is insignificant when 
three oxides layers are formed. The same result was obtained even for the 
two oxides layers. Since our DOS curves were smoothed, there is effect of 
the presence of the states in band gap. Thus the obtained results allow us 
to confirm the assumption made in Ref. \onlinecite{Vogtenhuber1994} that the properties of TiO$_{2}$ 
layers will be important in order to describe the interactions of TiNi 
implants with different tissue from first principles. 

To summarize, we have investigated the surface electronic structure and its 
change upon MT for two SME alloys using FLAPW method. The obtained results 
allow us to draw the conclusion that the $B$2 TiNi (110) surface is most 
preferable for martensitic reconstruction and surface relaxation can 
stimulate the martensitic transformations. Nevertheless only surface cannot 
be the only driving force for MT but the surface phenomena can influence MT 
with respect to martensitic temperature. The coherent boundary between 
austenite-martensite may be also the surface free energy. This fact 
together with observed ES changes in the $B$19' (010) surface allow us to 
conclude that growth of TiNi crystal will be preferable in this plane. MT 
does not influence considerable the surface states and their interaction 
with different adsorbates also, it is believed. The obtained ES of 
investigated surfaces in both austenitic and martensitic TiNi and TiPd 
phases showed that Ti atoms are highly reactive. This explains the existence 
of titanium oxides film at the TiNi surface. The simulation of oxides film 
growth on the TiNi (110) substrate was performed. 

\section*{ACKNOWLEDGMENTS}

This work was jointly supported by the Russian Foundation for Basic Research 
(grant N 02-02-16336), by the Korea Science and Engineering Foundation 
(KOSEF-2002) and by POSTECH basic research fund.

\clearpage
\begin{figure}[t]
\begin{center}
\includegraphics[keepaspectratio,width=0.3\columnwidth]{fig1a.eps}
\includegraphics[keepaspectratio,width=0.3\columnwidth]{fig1b.eps}
\includegraphics[keepaspectratio,width=0.3\columnwidth]{fig1c.eps}
\\
a)\hspace{5.0cm}b)\hspace{5.0cm}c)
\end{center}
\caption{
Atomic structure of the $B$19'-TiNi (100), (010) and (001) surfaces,
respectively (side view). Upper part of slabs used for calculations is
given only.}
\label{fig1}
\end{figure}

\clearpage
\begin{figure}[t]
\includegraphics[keepaspectratio,width=0.5\columnwidth]{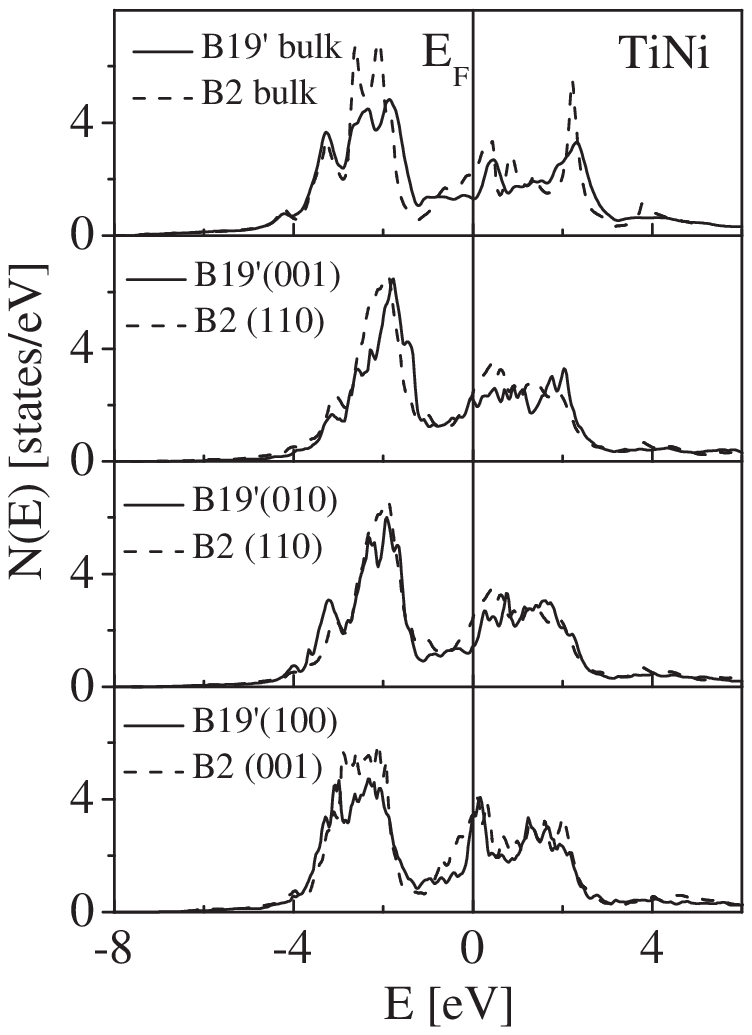}
\includegraphics[keepaspectratio,width=0.5\columnwidth]{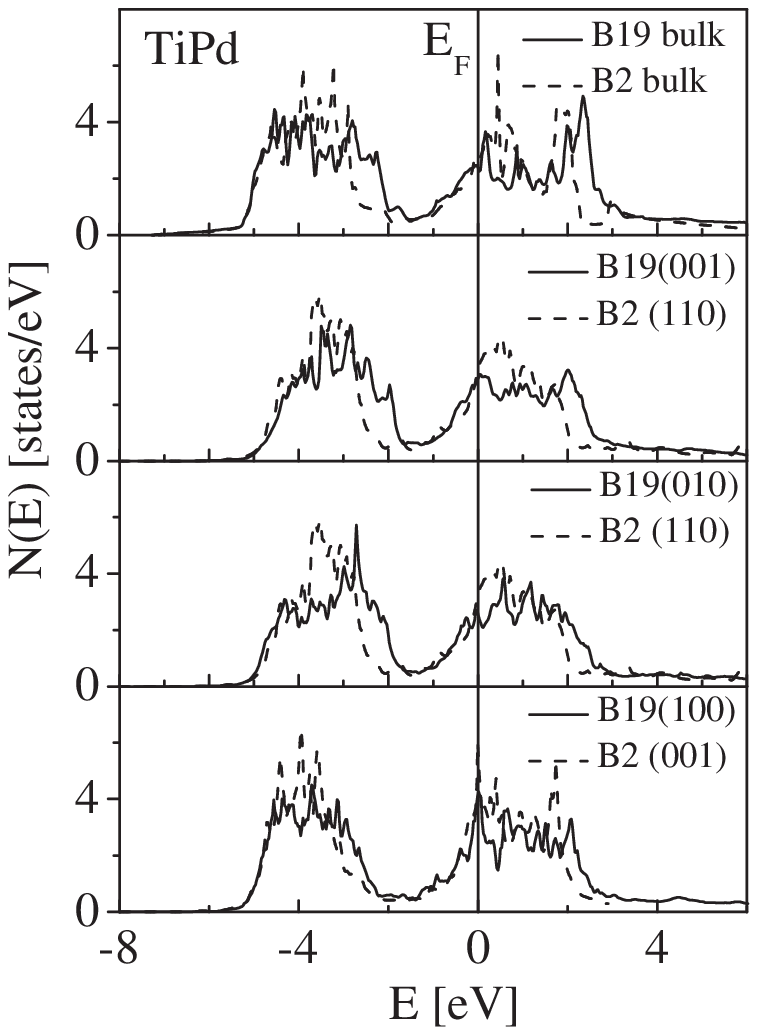}
\vspace{-1.5cm}
\begin{center}
a)\hspace{8.0cm}b)
\end{center}
\caption{
The surface ES of $B$19'-TiNi and $B$19-TiPd (solid lines) in
comparison with corresponding ones for parent $B$2-phase (dashed
lines).}
\label{fig2}
\end{figure}

\clearpage
\begin{figure}[t]
\includegraphics[width=0.5\columnwidth]{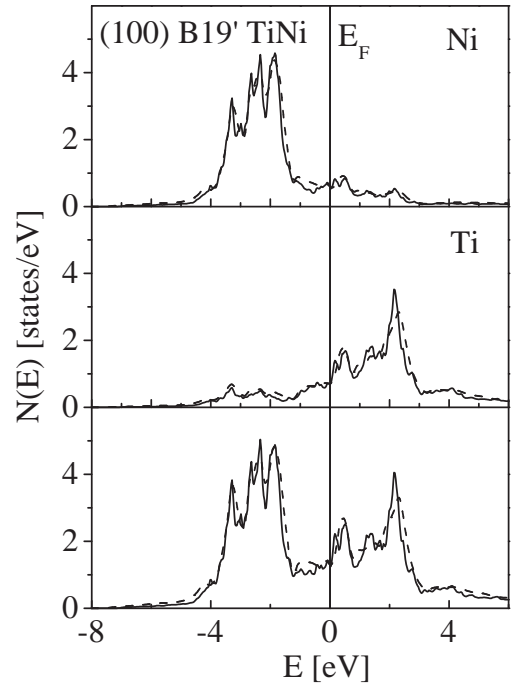}
\includegraphics[width=0.5\columnwidth]{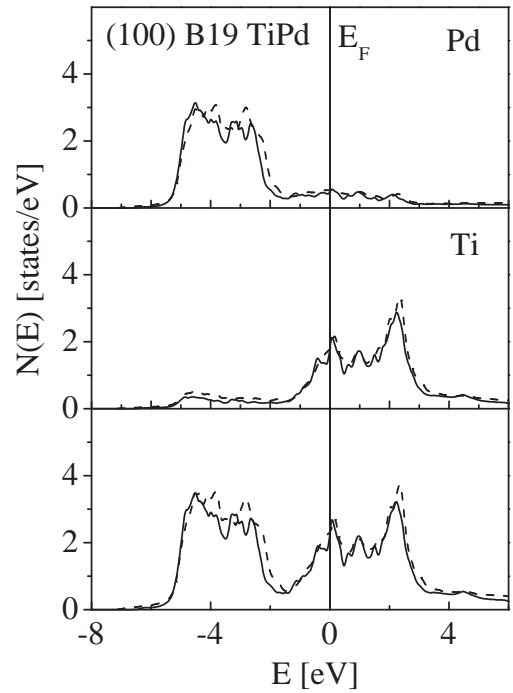}
\caption{
The calculated LDOS of central layers and their sum for $B$19'-TiNi
(100) surface (solid lines) in comparison with DOS for bulk alloy
(dashed lines).}
\label{fig3}
\end{figure}

\clearpage
\begin{figure}[t]
\includegraphics[width=0.5\columnwidth]{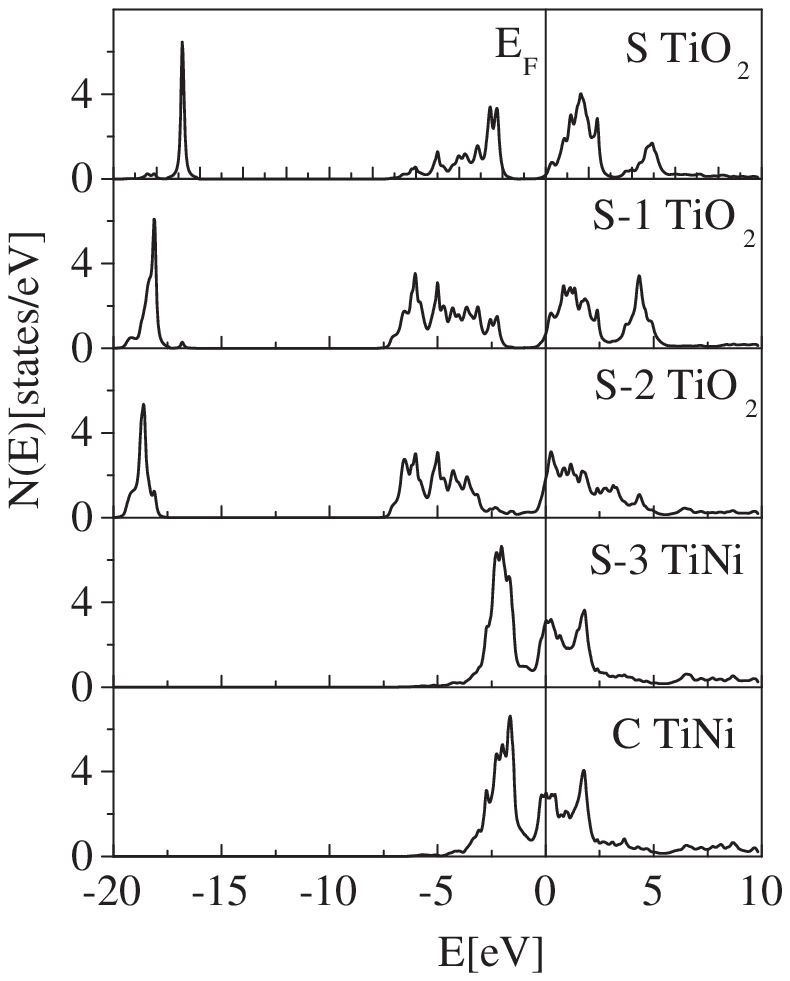}
\includegraphics[width=0.5\columnwidth]{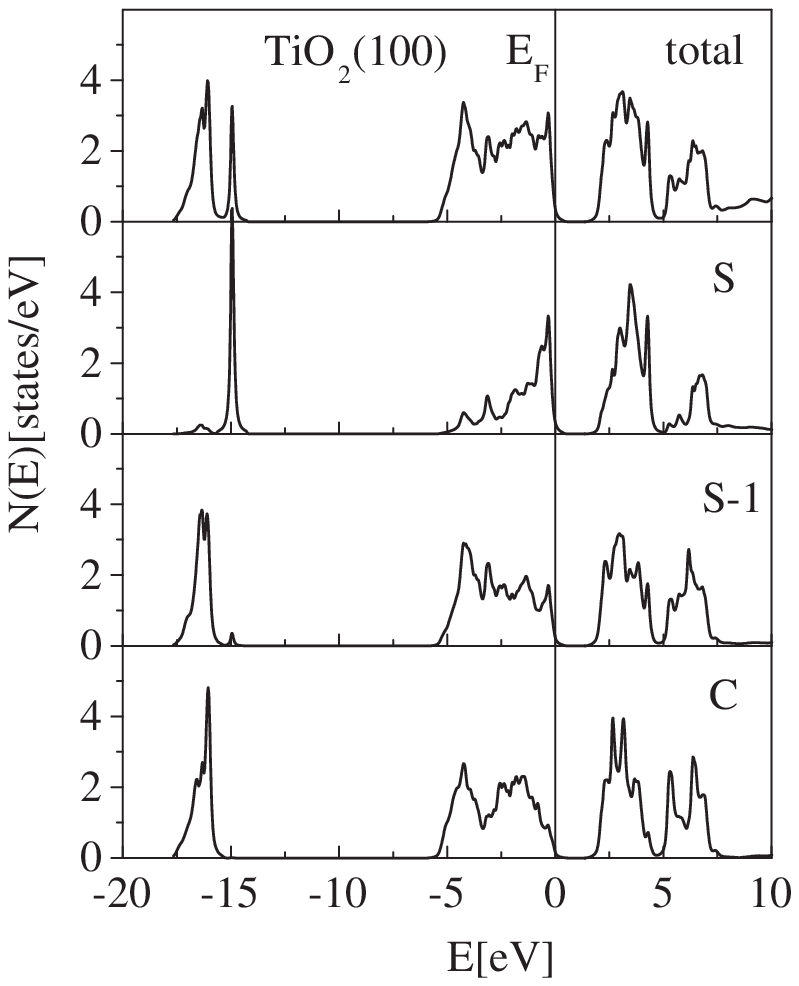}
\vspace{-1.5cm}
\begin{center}
a)\hspace{8.0cm}b)
\end{center}
\caption{
The layer-resolved DOS for TiO$_2$/TiNi system in comparison with
results for clean TiO$_2$ (100) surface. The symbols S and C mark the
surface and central layers, S-1, S-2, S-3 correspond to the position
from the surface.}
\label{fig4}
\end{figure}

\end{document}